\documentclass[reprint,amsmath,amssymb,aps]{revtex4-2}

\usepackage{graphicx}
\usepackage{dcolumn}
\usepackage{bm}

\begin{document}

\preprint{APS/123-QED}

\title{Ruling Out New Physics at Low Redshift as a solution to the $H_0$ Tension}

\author{Ryan E. Keeley}
\email{rkeeley@ucmerced.edu}
\affiliation{Department of Physics, University of California Merced, 5200 North Lake Road, Merced, CA 95343, USA}
\author{Arman Shafieloo}
 \email{shafieloo@kasi.re.kr}
\affiliation{Korea Astronomy and Space Science Institute (KASI), 776 Daedeok-daero, Yuseong-gu, Daejeon 34055, Korea}
\affiliation{KASI Campus, University of Science and Technology, 217 Gajeong-ro, Yuseong-gu, Daejeon 34113, Korea}

\date{\today}

\begin{abstract} 
We make the case that there can be no low-redshift solution to the $H_0$ tension. To robustly answer this question, we use a very flexible parameterization for the dark energy equation of state such that every cosmological distance still allowed by data exists within this prior volume. To then answer whether there exists a satisfactory solution to the $H_0$ tension within this comprehensive parameterization, we constrained the parametric form using different partitions of the Planck cosmic microwave background, SDSS-IV/eBOSS DR16 baryon acoustic oscillation, and Pantheon supernova datasets. When constrained by just the cosmic microwave background dataset, there exists a set of equations of state which yields high $H_0$ values, but these equations of state are ruled out by the combination of the supernova and baryon acoustic oscillation datasets. In other words, the constraint from the cosmic microwave background, baryon acoustic oscillation, and supernova datasets together does not allow for high $H_0$ values and converges around an equation of state consistent with a cosmological constant. Thus, since this very flexible parameterization does not offer a solution to the $H_0$ tension, there can be no solution to the $H_0$ tension that adds physics at only low redshifts. This is directly related to the expansion history of the Universe and its geometrical properties and would include models beyond those parametrized by $w(z)$.
\end{abstract}

\maketitle

\section{Introduction}
Finally with the precision era of cosmology we can test the theoretical underpinnings of $\Lambda$CDM, the concordance model of cosmology. That multiple different probes and datasets yield percent level constraints on various cosmological parameters have prompted the community to check the most basic astrophysical and theoretical underpinnings of these cosmological measurements.  Indeed, the emergence of the often discussed ``$H_0$ tension'' has reached a level of precision that demands some change in our understanding of the Universe. 

The $H_0$ tension is a mismatch in the value of the present Hubble rate ($H_0$) as directly measured by the Cepheid calibration of supernova (SN) distances~\cite{Riess:2016jrr,Riess:2019cxk,Riess:2020fzl}, and the value as inferred from a $\Lambda$CDM-based inference of the cosmic microwave background (CMB)~\cite{Planck:2018vyg}.  This mismatch has reached a statistical significance of over $5\sigma$~\cite{Riess:2021jrx}.  

As other methods of calibrating SN distances yield a much reduced tension with the CMB,~\cite{Freedman:2019jwv} a systematic error in the Cepheid calibration has not been ruled out. The question of systematics will have to wait until additional probes reach comparable precision~\cite{Birrer:2020tax,LIGOScientific:2017adf,LIGOScientific:2019zcs}. However, a new physics explanation is the more exciting possibility, as evidenced by the shear number of papers written that propose solutions. Though the number of proposed solutions is large, they can often be sorted into two (not mutually exclusive) categories, based on whether the new physics plays a role at high redshift or at low redshift~\cite{Linder:2021ujs,Abdalla:2022yfr}. By high and low redshift, we specifically mean pre and post-recombination, respectively.

Examples of physics at low redshift include curvature~\cite{Ryan:2019uor,DiValentino:2019qzk,2021PhRvD.103d1301H}, evolving dark energy~\cite{2018PhRvD..97l3501J,Addison:2017fdm,Lemos:2018smw,Aylor:2018drw,2019JCAP...12..035K,Li:2019yem,Arendse:2019hev,2020PhRvD.101j3517B}, dark matter - dark energy interactions~\cite{DiValentino:2020kpf} among others~\cite{DiValentino:2020kha}. Examples of physics at high redshift include extra radiation~\cite{2016JCAP...10..019B}, early dark energy~\cite{Poulin:2018cxd,Sakstein:2019fmf,Hill:2020osr}, interacting neutrinos~\cite{Kreisch:2019yzn,Park:2019ibn} and features in the primordial power spectrum~\cite{Hazra:2018opk,2020JCAP...09..055K,Hazra:2022rdl,Antony:2022ert}, among others. Two of many (see~\cite{Schoneberg:2021qvd,Abdalla:2022yfr} and references therein) such models that have been proposed to solve the $H_0$ tension are the Transitional Dark Energy (TDE) model~\cite{2019JCAP...12..035K} and the Phenomenologically Emergent Dark Energy (PEDE) model~\cite{Li:2019yem}. These models are illustrative examples for what features a model that new low-redshift physics needs to have to compose a satisfactory solution to the $H_0$ tension and why they ultimately fail.

There are a few a priori compelling reasons to look to physics at low redshift for a solution to the ``$H_0$ tension''.  Indeed, since the $\Lambda$CDM-based inference of $H_0$ from the CMB is a projection in over three orders of magnitude in the scale factor, and since the $\Lambda$ part of $\Lambda$CDM is theoretically a mystery, one might even expect some sort of tensions to arise. There is also a practical utility to first looking at low-redshift physics for a solution. For example, 
the predictions of $\Lambda$CDM are remarkably consistent with the observed angular power spectrum.  So if new physics were to only enter at low redshift then, such low-redshift new physics would only have to contend with purely geometric degeneracies in the CMB likelihoods,  e.g. making sure $\theta_s$ remains invariant, and thus the acoustic peaks of the CMB are not shifted~\cite{2018PhRvD..97l3501J,Knox:2019rjx}.

The guard rails of the tension, the SN and baryon acoustic oscillation (BAO) datasets, do not constrain $H_0$ on their own but map out the shape of the expansion history between $z=0$ and $z\sim 2.5$ and so are crucial for testing any low-redshift solution~\cite{Keeley:2020aym}.

Previous works, such as \citet{Camarena:2021jlr}~and ~\citet{Efstathiou:2021ocp}, have shown that a confusion about what physics is inferred if one uses the SH0ES information as a prior on $H_0$ or a prior on $M_B$ can disfavor specific classes of very low-redshift solutions to the $H_0$ tension.  In contrast and complimentary to these works, we seek to be more general and rule out all potential low-redshift solutions to the $H_0$ tension by whether or not they fit the entirety of cosmological datasets.

Despite, these a priori motivations for low-redshift explanations for the $H_0 $ tension, in this paper we will demonstrate that low-redshift solutions as a class are insufficient to provide a satisfactory solution to the $H_0$ tension. To do this we will employ a broad, flexible parameterization of the dark energy equation of state ($w(z)$) that would bracket every potential low-redshift solution.  That is every set of cosmological distances that are still allowed by the data would correspond to a $w(z)$ within this parameterization.  Thus, if there is no satisfactory solution to the $H_0$ tension within this parameterization, then there is no satisfactory low-redshift solution. We will additionally discuss intuitively why these results should be expected based on low-redshift solutions' inability to satisfy a number of ``tension triangles''.

\section{Datasets}
There are four key datasets that we use in this analysis, the Planck CMB datset~\cite{Planck:2018vyg},the SDSS-IV/eBOSS BAO dataset~\cite{eBOSS:2020yzd}, the Pantheon SN dataset~\cite{Pan-STARRS1:2017jku}, and we compare results from these datasets with the SH0ES $H_0$~\cite{Riess:2021jrx} constraint.

The SH0ES dataset~\cite{Riess:2021jrx} includes observations of 42 Cepheids in host galaxies of SN at $z<0.01$.  The Cepheids, whose distances are inferred from a period-luminosity relationship, are in turn calibrated using geometric parallaxes from \textit{Gaia} EDR3, masers in NGC 4258 and detached eclipsing binaries in the Large Magellanic Cloud. Using calibrated Cepheids to measure the distance to local SN amounts to a calibration of the SN luminosity, often parametrized by the B-band absolute magnitude $M_B$, which in turn amounts to a constraint on $H_0 = 73.04 \pm 1.04$.

Type Ia SN are useful for measuring cosmological distances since they are empirically assumed to be standardizable candles. That is, SN are thought to have the same intrinsic luminosity once various properties of the SNs' light-curves (e.g the stretch and color) are properly calibrated as well as properties of the host galaxy.
Since this intrinsic luminosity, parameterized by $M_b$, is unknown, and since it is degenerate with $H_0$, one needs an independent anchor like Cepheids to complete the calibration.  Thus, on their own, measuring the relative brightness of a compilation of SN can yield information about only their relative distances and cannot measure $H_0$. The publicly-released Pantheon SN dataset is composed of 1048 Type Ia SN between $z = 0.01$ and $z=2.3$~\cite{Pan-STARRS1:2017jku} and the calibration of the SN light-curve parameters (namely not $M_b$) has already been performed and the corresponding uncertainties are included as a systematic component to the total covariance matrix. $M_b$ is varied in our analysis whenever we include this dataset.

The SDSS-IV/eBOSS DR16 baryon acoustic oscillation (BAO) dataset measures the positions and redshifts of galaxies and uses such catalogs to reconstruct the correlation function of galaxies~\cite{eBOSS:2020yzd}. 
This correlation function contains a ``BAO feature'' which is an overdensity of power at the drag scale $r_s$. 
Since the true distances to these galaxies are unknown, the reconstruction of the position of the BAO feature has to be done in a dimensionless, unanchored space, where the peak is found at a position of $\sim 100 h^{-1}$ Mpc. Thus, using these reconstructed correlation functions to measure cosmological distances returns constraints on the Hubble distance, $D_H(z) = c/H(z)$, and the angular diameter distance $D_M(z)$, both relative to $r_s$.

The Planck satellite ~\cite{Planck:2018vyg} measures the anisotropies in the temperature and polarization of the CMB.  We use the `TT', `TE' and `EE' parts of the Planck 2018 dataset. In the context of analyzing solutions to the $H_0$ tension that only include new physics at low redshift and correspondingly have identical physics to $\Lambda$CDM at high redshift, the CMB primarily constrains any new low-redshift physics via geometric degeneracies~\cite{2019JCAP...12..035K}.  For instance, changing $H_0$ induces a phase-shift in the CMB's acoustic peaks~\cite{2020JCAP...09..055K}.  In such a case, the information contained in the CMB can be summarized as constraints in the Hubble parameter at and angular diameter distance to the surface of last scattering $H(z_*)$ and  $D_A(z_*)$.

\section{Results}

\begin{figure}
    \centering
    \includegraphics[width=\columnwidth]{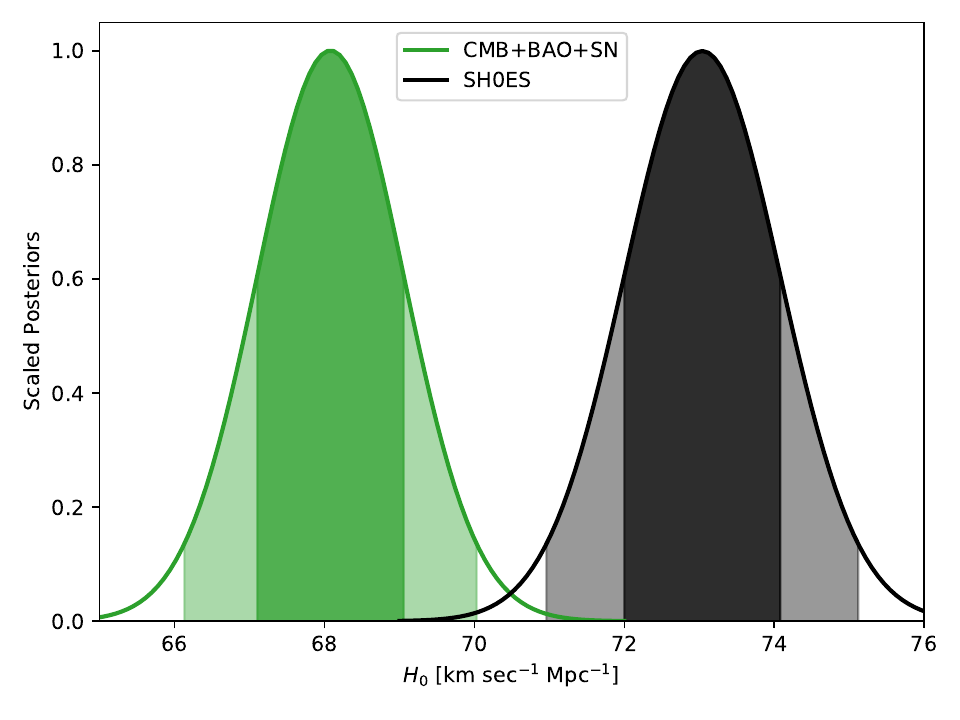}
    \caption{1D posterior of $H_0$ from the joint CMB+BAO+SN constrain on the Chebyshev parameterization, along with the SH0ES result for comparison.}
    \label{fig:H0}
\end{figure}

\begin{figure}
    \centering
    \includegraphics[width=\columnwidth]{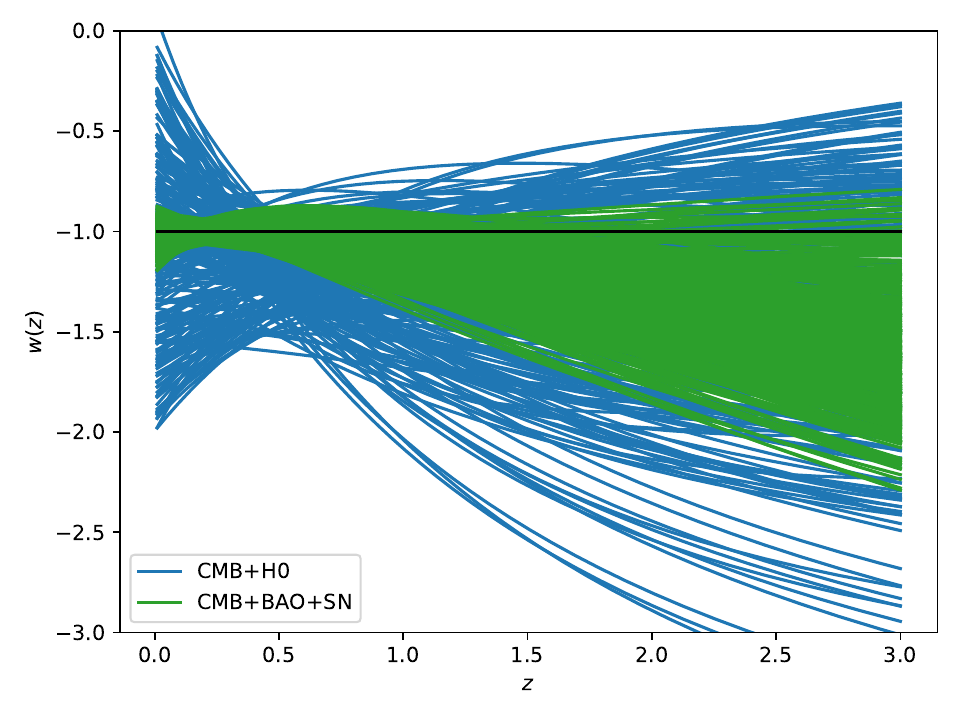}
    \caption{PPD for $w(z)$ from the CMB+H0 constraint in blue and from the CMB+BAO+SN constraint in green. The black line indicates $w(z)=-1$, the $\Lambda$CDM equation of state.}
    \label{fig:w_z}
\end{figure}

In the appendix, we have demonstrated that the Chebyshev parameterization (Crossing function) should include any potential low-redshift solution to the $H_0$ tension should one exist, we now test whether one does actually exist. To do this, we constrain the Chebyshev parameterization using the CMB+BAO+SN datasets and compute the posterior of $H_0$.  If the posterior from the joint constraint from the CMB+BAO+SN datasets spans the $H_0$ constraint from SH0ES, then, with this model, the joint CMB+BAO+SN datasets would not be in tension with the SH0ES dataset.  Then we would properly calculate a joint CMB+BAO+SN+H0 constraint and conclude that the Chebyshev parameterization offers a satisfactory solution to the $H_0$ tension. Conversely, if the constraint on $H_0$ from the joint CMB+BAO+SN datasets is still in tension, then we do not calculate a joint constraint with the SH0ES $H_0$ constraint as the results would be meaningless. The results of this test are shown in green in Fig.~\ref{fig:prior}. Unfortunately, we can see in that figure that the SH0ES constraint lies entirely outside posterior predictive distributions for the Chebyshev model using the CMB+BAO+SN dataset.  Indeed, we find that $H_0 = 68.08 \pm 0.97$ km sec$^{-1}$ Mpc$^{-1}$~(see~\ref{fig:H0}), which relaxes the tension but does not offer a satisfying solution.  Thus the Chebyshev parameterization does not solve the $H_0$ tension, and by extension, no instance of new physics that only plays a role at low redshift would solve the $H_0$ tension. 

In Fig.~\ref{fig:w_z} we see another way to show both the flexibility of the Chebyshev parameterization and the fact that the CMB+BAO+SN datasets constrain it well enough to disfavor any new physics beyond $w(z)=-1$ as a solution to the $H_0$ tension.  The blue curves in Fig.~\ref{fig:w_z} are $w(z)$ functions drawn from the posterior of the Chebyshev parameterization constrained by the CMB+H0 dataset.  The takeaway from this distribution is the variety of $w(z)$ functions that can achieve a $H_0$ value consistent with the SH0ES constraint.  The green curves are functions from the posterior of the Chebyshev parametrziation constrained by the CMB+BAO+SN dataset. That these curves converge around $w(z)=-1$ shows that the CMB+BAO+SN dataset shows no preference for any deviation fro $\Lambda$CDM.  There is a lot of flexibility still allowed in the phantom regime especially above $z>1$.  This is, after some thought a rather obvious point.  There are two effects, above $z>1$ the data become less constraining for both datasets so one can observe an increased scatter in $D_H(z)$ and $D_M(z)$, but also, above $z>1$ the dark energy is becoming less of the dominant component of the expansion rate, and so the dynamics of its equation of state is less relevant for the fit.  Indeed, for the same dynamics of $w(z=0)$, $w(z=2)$ can vary from $w(z=2) = -1$ to $w(z=2) = -2$ with only a change in the log-likelihood of $\sim 0.1$.

\section{Discussion}

In this section we discuss why low-redshift solutions to the $H_0$ solution are insufficient and seek to develop intuition about what features a successful solution to the $H_0$ tension would satisfy. 

\begin{figure}
    \centering
    \includegraphics[width=\columnwidth]{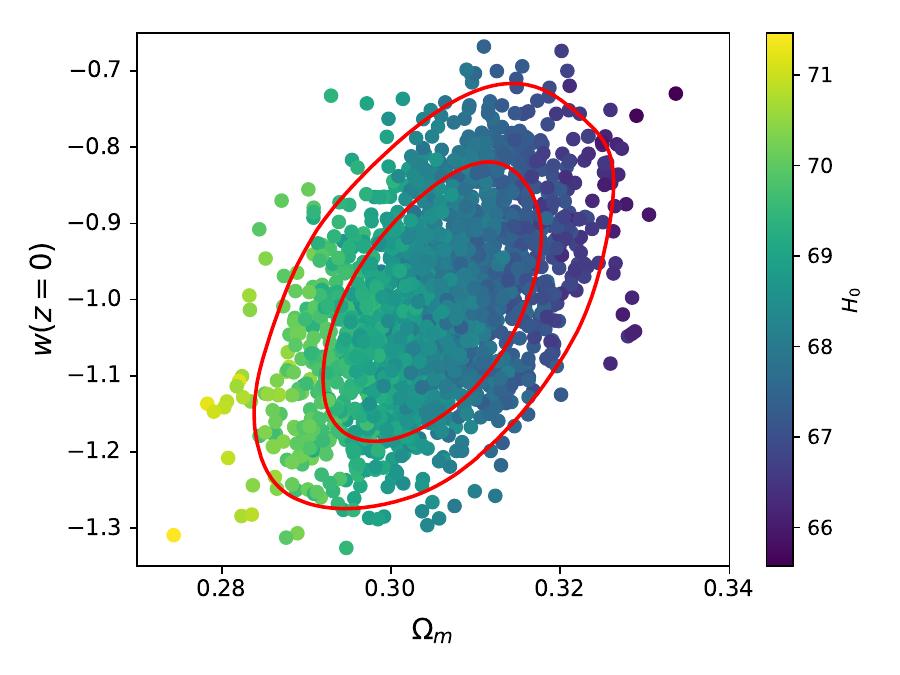}
    \caption{Constraints on ($w(z=0)$) and $\Omega_{\rm m}$ from the Chebyshev model using the CMB+SN+BAO datasets. The red contours correspond to the $1$ and $2$-$\sigma$ confidence levels.  The individual points are samples from the MCMC chain and the color of each point corresponds to the $H_0$ value. The color gradient demonstrates the correlation between the parameters of the shape of the Hubble function and that the SH0ES-preferred $H_0$ values are beyond the region allowed by the other datasets.}
    \label{fig:correlations}
\end{figure}

Any proposed solution to the $H_0$ tension is a prediction that the proposed extended parameters are correlated with a high $H_0$ value.  In Fig.~\ref{fig:correlations}, we see how the extended parameters of the Chebyshev model are correlated with the standard $\Lambda$CDM model. This figure shows the posterior for the Chebyshev model using the joint CMB+BAO+SN dataset. The Chebyshev parameterization is complicated but the important part of it when fitting the CMB+BAO+SN dataset is $w(z=0)$, the value of the equation of state that is most stringently constrained by the data (the SN data are very dense around $z\sim 0.1$). So for this parameterization, a high $H_0$ value requires $w(z=0)<-1$, and at the same time a high $H_0$ value requires a low $\Omega_m$ value to satisfy the constraint on $\Omega_m h^2$ from the CMB.  Especially taken together, a low $w(z=0)$ value and a low $\Omega_m$ value make a poor fit to the SN data which drive the fit towards $w(z=0)=-1$ and $\Omega_m\sim0.3$. So its ultimately the SN constraint that rules out much of the Chebyshev parameterization's extended parameter space.

\begin{figure}
    \centering
    \includegraphics[width=\columnwidth]{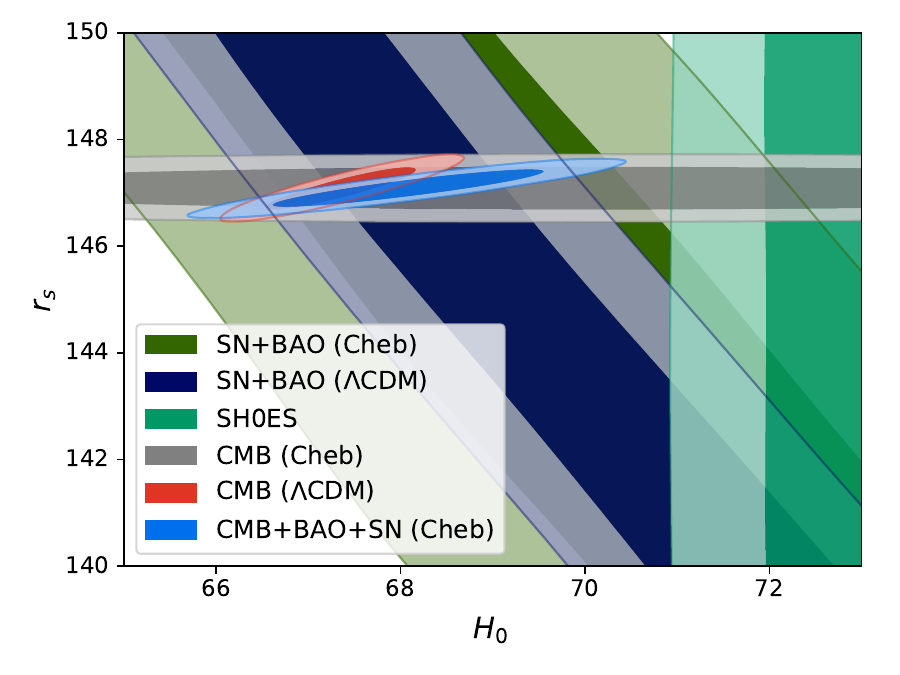}
    \caption{1 and 2$\sigma$ constraints on $H_0-r_s$ for the $\Lambda$CDM and Chebyshev models using the CMB (red and grey, respectively), BAO+SN (navy and olive), and CMB+BAO+SN (blue) datasets. For reference, we include the SH0ES constraint in green.}
    \label{fig:H0rs}
\end{figure}

Shown in Fig.~\ref{fig:H0rs} are constraints for the Chebyshev model using the CMB (grey), the BAO+SN (olive), and CMB+BAO+SN (blue) and for the $\Lambda$CDM model using the CMB (red) and BAO+SN (navy).
One might look at this triangle plot and think there exists a point in parameter space where the grey overlaps the olive which overlaps the green and thus there would exist a solution to the $H_0$ tension.  However, these plots are a projection of a large dimensional posterior onto two dimensions and in this higher dimensional posterior, there is no overlap.  The beyond-$\Lambda$CDM parameters that the Planck posterior towards SH0ES constraint are different from the ones that expand the BAO+SN posterior.  For the $\Lambda$CDM constraints, the CMB, BAO+SN, and SH0ES constraints create a ``tension triangle''~\cite{2016JCAP...10..019B}, which results from the three constraints never overlapping at any one point. This tension triangle evidences the need to modify $r_s$ as part of solving the $H_0$ tension. That the Chebyshev constraints on the CMB and BAO+SN datasets expand towards the SH0ES constraint, one might conclude the modifying $r_s$ is, in fact, not needed. Indeed, the Chebyshev parameterization breaks the degeneracy between $r_s$ and $H_0$.  However, the full CMB+BAO+SN constraint for the Chebyshev model shrinks away from the SH0ES constraint, indicating the expanded parameter space does not alleviate the need for modifying $r_s$.

\begin{figure}
    \centering
    \includegraphics[width=\columnwidth]{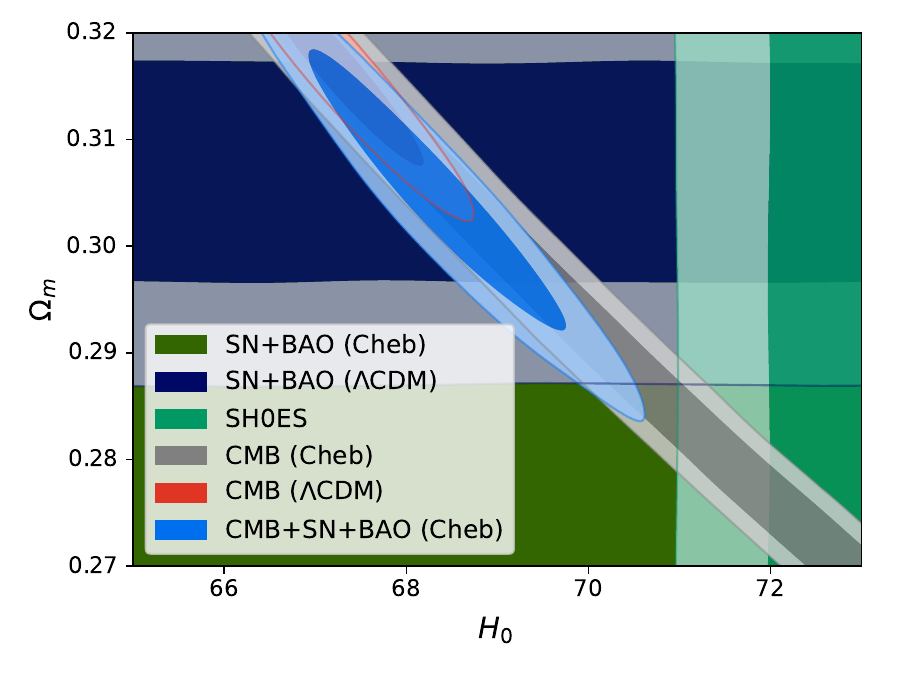}
    \caption{As Fig.~\ref{fig:H0rs} but for $H_0-\Omega_m$.}
    \label{fig:H0OmTri}
\end{figure}

As with Fig.~\ref{fig:H0rs}, Fig.~\ref{fig:H0OmTri} shows that the CMB, BAO+SN, and SH0ES constraints create a tension triangle also in the parameters $H_0$ and $\Omega_m$. Such a tension triangle has been advanced as a reason why simply modifying $r_s$ cannot fully resolve the $H_0$ tension~\cite{2021CmPhy...4..123J}.
From the CMB, the parameter $\Omega_m h^2$ can be measured independently of low-redshift physics.  This fact is why Chebyshev constraint from the CMB lies along the line of degeneracy $\Omega_m h^2$ being a constant. The flexible Chebyshev parameterization breaks any degeneracy between $\Omega_m h^2$ and $H_0$ that exists in the $\Lambda$CDM model.  
A similar story unfolds between $H_0$ and $\Omega_m$ as with $H_0$ and $r_s$.  Between the $\Lambda$CDM fits there is no parameter where the CMB, BAO+SN, and SH0ES constraints overlap. Even relaxing the CMB constraint via the Chebyshev model only creates a narrow overlapping region at the edge of the 2$\sigma$ region of each of the constraints. When BAO+SN constraint is relaxed via the Chebyshev model, there is a large region where the three constraints overlap, however, when we look at the Chebyshev CMB+BAO+SN constraint, it does not overlap with the SH0ES constraint, because the region of the extended parameter space that allowed the CMB constraint to expand towards larger $H_0$ values is different than the region that allowed the BAO+SN constraint to expand towards lower $\Omega_m$ values.

To fit both the $H_0$ and CMB constraints, a successful model that adds physics at low redshift must have a faster-than-$\Lambda$CDM expansion history at low redshift and a slower-than-$\Lambda$CDM expansion history at high redshift.  The simplest instance of this feature is a purely phantom $w(z)$.  To further fit the SN well, the model would then need to account for the fact the CMB+H0 constraint is pulling $\Omega_m$ to lower values.  To counteract this effect, the $w(z)$ can transition to a quintessent regime, particularly the transition is constrained to be $z>1$, the region where the SN were less constraining and the BAO constraint did not extend to at the time.  Such a transition from a phantom to a quintessent regime is what gave the TDE model it's name. However, for models like TDE, it is ultimately the BAO constraints that the model fails to explain. Particularly the high-redshift ($z>1$) BAO measurements constrain $H(z)$ to be around $\Lambda$CDM values where the transition was previously happening. Those BAO measurements rule out the final slice of parameter space that allowed the TDE model to work as a satisfactory solution to the $H_0$ tension.

\section{Conclusions}
Even with a very flexible model, such as the Chebyshev parameterization including curvature, that brackets the entirety of the relevant space of model uncertainties and can fit every relevant cosmological distance, there exists no point in the extended parameter space that adequately explains the $H_0$ tension.  Thus, since this model brackets the entirety of the low-redshift space of model uncertainties, its inability to find a solution to the $H_0$ tension indicates that there is no satisfactory solution to the $H_0$ tension that adds physics at only low redshift. This argument holds even for models of low-redshift physics that are not explicitly parameterized by $w(z)$ such as DM-DE interactions or MG. Thus, the community should look for high-redshift modifications to the standard $\Lambda$CDM model, with its power-law form of the primordial power spectrum, to provide a satisfying new physics solution to the $H_0$ tension. 

\section*{Acknowledgements}
We would like to thank Eric Linder for useful comments on the draft. This work was supported by the high performance computing cluster Seondeok at the Korea Astronomy and Space Science Institute. A.~S. would like to acknowledge the support by National Research Foundation of Korea NRF-2021M3F7A1082053 and Korea Institute for Advanced Study (KIAS) grant funded by the government of Korea.

\appendix

\section{Chebyshev Model and Crossing statistics}\label{sec:app}
\begin{figure}
    \centering
    \includegraphics[width=\columnwidth]{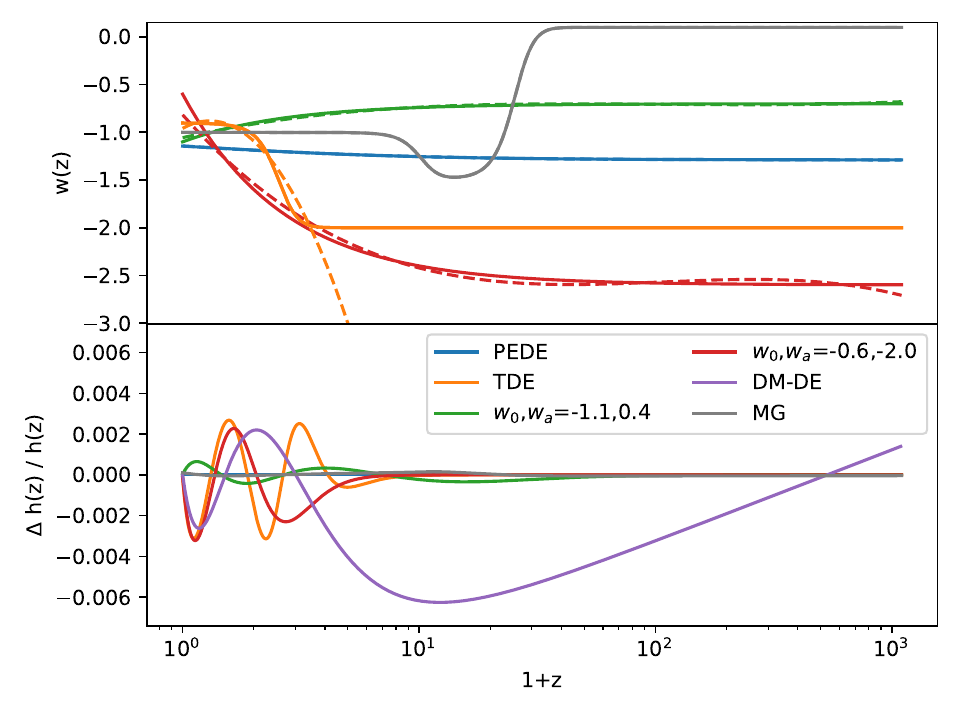}
    \caption{(Top panel) Example equation of states that have previously been proposed to solve the $H_0$ tension (solid lines), the TDE model ~\cite{2019JCAP...12..035K}, the PEDE model~\cite{Li:2019yem}, and an example MG model~\cite{Kase:2018iwp}, along with CPL parameters within the $2\sigma$ region of joint Pantheon+ and Planck constraint, as well as the Chebyshev polynomial that best matches (dashed lines) those models. (Bottom panel) Fractional difference in $h(z)$ between the example equation of states and the Chebyshev parameterization that most closely matches them. The bottom panel also includes an example DM-DE model~\cite{DiValentino:2017iww, Yang:2018euj}.}
    \label{fig:examples}
\end{figure}

Since the datasets in question are primarily geometrical, we only need to parametrize the Friedmann equation in order to explain them,
\begin{equation}
    h^2(z) = \Omega_m (1+z)^3 + \Omega_k (1+z)^2 + \Omega_{DE}(z) + \Omega_r(z)  ,
\end{equation}
where $h(z) = H(z)/H_0$, $\Omega_m$ is the matter density, $\Omega_k$ is the curvature, $\Omega_{r}(z)$ is the radiation density which is fixed to be the fiducial Planck $\Lambda$CDM values, 
and $\Omega_{DE}(z)$ is the dark energy density. An evolving dark energy density is related to the dark energy equation of state $w(z)$, thusly,
\begin{equation}
    \Omega_{DE}(z) = (1-\Omega_m-\Omega_k-\Omega_r(0))\exp\left(3 \int dz \frac{1+w(z)}{1+z}\right). 
\end{equation}
We are implicitly assuming that even though the dark energy density evolves, the dark energy does not contribute to clustering, and the sound speed of dark energy can be set such that is a valid approximation.
Thus the new physics that we are testing at low redshift amounts to a $w(z)$ different from $-1$, specifically
\begin{equation}
    w(z) = -\sum_0^3 c_i T_i(x), \ x = \log(1+z)/\log(1+z_*).
\end{equation}
where $T_i(x)$ are the the Chebyshev polynomials and $c_i$ are coefficients that are varied in our analysis. We model $w(z)$ up to $z_*$, the redshift of the surface of last scattering, in order to model any potential new physics that plays a role after recombination. We could use higher order Chebyshev polynomials but we find four are sufficiently flexible. Previously, Chebyshev polynomials have been used in the context of the Crossing statistic, a technique used to check the internal consistency of a dataset as well as searching for deviation from a theoretical model or a parametric form~\cite{Shafieloo:2012jb,Shafieloo:2012yh,Shafieloo:2016zga}. In this work, Chebyshev polynomials act as Crossing functions to go beyond the flexibility of the standard $\Lambda$CDM model fitting combinations of cosmological observations. In total, we vary the standard 6 parameters of the $\Lambda$CDM model ($\theta_s$, $\Omega_bh^2$, $\Omega_ch^2$, $\tau$, $A_s$, $n_s$)
, curvature $\Omega_k$, as well as each of $c_{1,2,3,4}$. Further, we also vary the nuisance parameter $M_b$ when we include the SN likelihood. We adopt flat priors on each of these parameters.


There are two crucial properties that we need to consider to have a parametric form that can be inclusive enough to cover various forms of the evolving dark energy models. These two are orthogonality and convergence within the limited range.  Orthogonality allows us to cover a wide range of behaviors with minimum number of coefficients (degrees of freedom) and the convergence within a limited range is crucial for us so we can use the functional form to fit cosmology data that has a clear redshift range. It is true that other orthogonal polynomials can be used here but they should also satisfy the property of convergence in a limited range. While one can use the Gram-Schmidt process to generate a set of orthogonal bases tailored appropriately for the case being studied, Chebyshev polynomials naturally have these crucial characteristics that allow us to use them trivially. Further supporting the reason we used Chebyshev polynomials in this work is that we have found them to be an efficient and precise basis for reconstruction of the Universe’s expansion history in past works~\cite{Shafieloo:2012jb,Shafieloo:2012yh,Shafieloo:2016zga}.

In general, introducing a large number of extra model parameters is not recommended in the context of model selection and parameter estimation, since interpreting the 2D joint posteriors of the extra model parameters can be complicated. Indeed, testing whether these extra parameters are statistically significant by visually inspecting if parameters corresponding to the base model are beyond the joint 2D posteriors of the extended parameters (e.g. for the Chebyshev parameterization, {$c_1$, $c_2$, $c_3$, $c_4$} = {1,0,0,0} corresponds to $\Lambda$CDM) can be misleading due projection effects.  For instance, {$c_1$, $c_2$, $c_3$, $c_4$} = {1,0,0,0} can be outside the full N-dimensional posterior but not in any of the joint 2D posteriors, and vice versa. This has been the case for the majority of the solutions proposed to resolve the Hubble tension. Since, in this analysis, we only care about whether this parameterization can model the high SH0ES $H_0$ value, we can avoid this sort of complication. Similarly, we plot the posterior distribution of the cosmological functions $D_H(z)$ and $D_M(z)$ for this reason. Finally, since having a large number of model degrees of freedom tends to greatly inflate confidence regions, that our parameterization still does not overlap the SH0ES constraint supports our argument that no low-redshift only solution can exist, despite the extra model degrees of freedom.  

In Fig.~\ref{fig:examples}, we show that our Chebyshev parameterization is flexible enough to approximate a variety of evolving dark energy models.  Two such models that we explicitly check are the TDE model~\cite{2019JCAP...12..035K} and the PEDE model~\cite{Li:2019yem}.  These two models are useful for making this point since they represent two opposite regimes of a slowly varying, purely phantom ($w<-1$) equation of state with the PEDE model and a quickly evolving equation of state that transitions between a phantom equation of state and a quintessent ($w>-1$) equation of state with the TDE model. Further, we use two examples from the Chevalier-Polarski-Linder (CPL) parameter space~\cite{Chevallier:2000qy,Linder:2002et} that are within the $2\sigma$ confidence region of the Pantheon+ and Planck joint constraint~\cite{Brout:2022vxf}.  We can see in Fig.~\ref{fig:examples} that the PEDE model equation of state can be matched, basically exactly, but the TDE equation of state is a less exact match. However, the corresponding $H(z)$ and $D_M(z)$ values are all within $<0.4$\%. There are other models that are not parameterized by $w(z)$ that can be well-approximated by our Chebyshev parameterization, where the data are constraining. For instance, $h(z)$ from the interacting dark matter and dark energy model (DM-DE)~\cite{DiValentino:2017iww, Yang:2018euj} is shown in Fig.~\ref{fig:examples}. Specifically, the parameters of the DM-DE model ($w_x$ and $\xi$) were constrained by the datasets considered as well as the Cheybshev parametrization. In the regimes where cosmological data are constraining, the Chebyshev parametrization can approximate the DM-DE model to within $<0.4\%$. Further, modified gravity (MG) models can be parametrized with the Chebyshev parametrization. For instance, there have been proposed beyond-Horndeski theories of gravity where the dark energy is a k-essence Lagrangian~\cite{Kase:2018iwp}. These attractor solutions can be matched exactly by our Chebyshev parametrization.   

\begin{figure}
    \centering
    \includegraphics[width=\columnwidth]{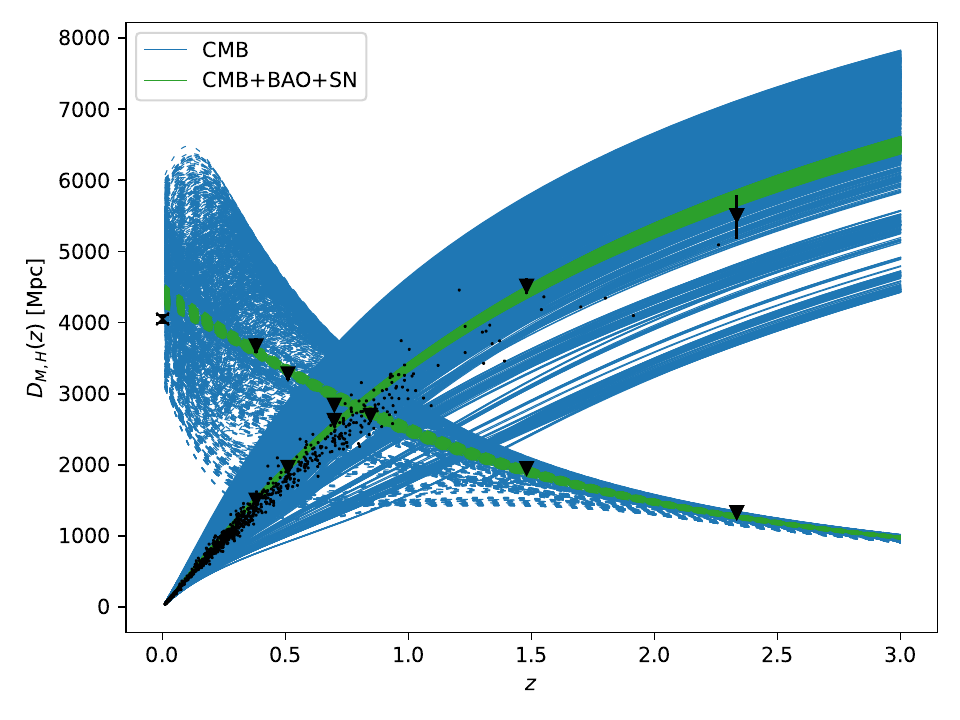}
    \caption{Posterior predictive distribution (PPD) for $D_H(z)$ (dashed, varies from upper-left to lower-right) and $D_M(z)$ (solid, varies from lower-left to upper-right) for the Chebyshev model using the CMB dataset alone (blue) and the CMB+BAO+SN. The black data points are from the SH0ES (``X''), Pantheon (dots), and SDSS/eBOSS datasets (triangles). The blue PPD demonstrates the model is flexible enough to include every cosmological distance that is still allowed by the data, while the green PPD demonstrates that, when constrained by the BAO and SN datasets, there is no longer enough flexibility to match the SH0ES constraint.}
    \label{fig:prior}
\end{figure}

In addition to comparing with specific models, another useful test to demonstrate the flexibility of the Chebyshev parameterization is whether it can bracket the data.  That is, whatever cosmological functions ($H(z)$, $D_M(z)$) that may still be allowed by the data would find a sufficiently close match in some location in the Chebyshev parameterization.  To this end, we calculate a posterior predictive distribution (the distribution of the cosmological functions $D_H(z) = c/H(z)$ and $D_M(z)$ that correspond to the posterior probability of the models' parameters) for the Chebyshev model constrained on just the Planck CMB data. The results of this calculation is shown in blue in Fig.~\ref{fig:prior}.  That the distribution spans the data indicates that this model is flexible enough to contain a low-redshift solution to the $H_0$ tension should one exist. 

\bibliography{apssamp}

\end{document}